\documentclass{article}

\usepackage{arxiv}

\usepackage{amsthm,amsmath}
\RequirePackage{hyperref}
\usepackage[utf8]{inputenc} 
\usepackage{smartdiagram} \usesmartdiagramlibrary{additions}
\usepackage[ruled]{algorithm2e}
\usepackage{dsfont}

\newcommand{\T}{^{\mathrm{T}}}

\title{Netboost: Boosting-supported network analysis improves high-dimensional omics prediction in acute myeloid leukemia and Huntington's disease}

\author{
  Pascal~Schlosser$^{1,2,*}$
  \And
  Jochen~Knaus$^2$
  \And
  Maximilian~Schmutz$^3$
  \And
  Konstanze~D\"ohner$^4$
    \And
  Christoph~Plass$^5$
   \And
  Lars~Bullinger$^6$
   \And
  Rainer~Claus$^3$
    \And
  Harald~Binder$^2$
  \And
  Michael~L\"ubbert$^{7,8}$
  \And
  Martin~Schumacher$^2$
}

\newcommand{\myitem}{\refstepcounter{enumi}\item[$^{*}$]}

\begin{document}
\maketitle

\begin{enumerate}
\item
Institute of Genetic Epidemiology, Faculty of Medicine and Medical Center - University of Freiburg, Freiburg im Breisgau, Germany

\item
Institute of Medical Biometry and Statistics, Faculty of Medicine and Medical Center - University of Freiburg, Freiburg im Breisgau, Germany

\item
Department of Hematology and Oncology, Augsburg University Medical Center, Augsburg,                              Germany

\item
Department of Internal Medicine III, University Hospital of Ulm, Ulm, Germany

\item
Division of Cancer Epigenomics, German Cancer Research Center, Heidelberg, Germany

\item
Hematology, Oncology and Tumor Immunology, Campus Virchow Hospital, Charite University Medicine, Berlin, Germany

\item
Department of Hematology-Oncology, Medical Center, Faculty of Medicine, University of Freiburg, Freiburg im Breisgau, Germany

\item
German Consortium for Translational Cancer Research (DKTK), Freiburg im Breisgau, Germany

\myitem
\texttt{pascal.schlosser@uniklinik-freiburg.de}
\end{enumerate}
\medskip

\begin{abstract} 
\textbf{Background:}\\
State-of-the art selection methods fail to identify weak but cumulative effects of features found in many high-dimensional omics datasets.
Nevertheless, these features play an important role in certain diseases.\\

\textbf{Results:}\\
We present Netboost, a three-step dimension reduction technique. First, a boosting-based filter
 is combined with the topological overlap measure to identify the essential edges of the network.
  Second, sparse hierarchical clustering is applied on the selected edges to identify modules and 
  finally module information is aggregated by the first principal components. The primary analysis
   is than carried out on these summary measures instead of the original data.\\
We demonstrate the application of the newly developed Netboost in combination with CoxBoost for
 survival prediction
of DNA methylation and gene expression data from 180 acute myeloid leukemia
(AML) patients and show, based on cross-validated prediction error
curve estimates, its prediction superiority over variable selection on the full dataset as well as over an
alternative clustering approach. The identified signature related to chromatin modifying enzymes was replicated in 
an independent dataset of AML patients in the phase II AMLSG 12-09 study.\\
In a second application we combine Netboost with Random Forest
classification and improve the disease classification error in RNA-sequencing data of Huntington's disease mice.\\

\textbf{Conclusion:}\\
Netboost improves definition of predictive variables for survival analysis and classification. It is a freely
available Bioconductor R package for dimension reduction and hypothesis generation in high-dimensional omics applications. 
\end{abstract}


\section*{Background}
Microarray, sequencing and other high-throughput functional genomics
technologies are developing rapidly, incorporating more and more
features. A major challenge in biomedical research is the
contrast of these high-dimensional datasets and the related investigation of potentially hundreds of thousands of features
 with only a limited sample population in the range of tens to a few hundreds. This gives rise to dimension reduction techniques with the
aim of identifying the subspace with minimal dimensions and the best
characterization of the outcome variable simultaneously\cite{Saeys2007}. 

Many times, methods which were originally developed for the selection of
tens of clinical variables are now faced with the challenge of
selecting from hundreds of thousands or even from millions of variables.
In cases where it is not expected that a singular feature dominates the effect but rather that a larger group of features works
cumulatively, the challenge becomes even greater.

In acute myeloid leukemia (AML) part of the epigenotype of
the disease is a global increase in DNA methylation in regulatory
regions \cite{Sobieszkoda2017}. Furthermore, for elderly patients the
only effective drugs that counteract this
effect are hypomethylating agents \cite{Gardin2017,Thomas2017, Papaemmanuil2016}. From this it is known that the state of
methylation fulfills an important role in this disease. Nevertheless
it has been difficult to incorporate DNA methylation markers in
patient relevant statistics like survival prediction \cite{Gardin2017,PradaArismendy2017}. Predictive methylation sites
in AML patients treated with chemotherapeutics \cite{yamazaki2016} and predictive sites from chronic
myelomonocytic leukemia patients treated with hypomethylating drugs
\cite{meldi2015} 
could not be
replicated for AML patients treated with hypomethylating
drugs.

Weighted Gene Co-expression Network Analysis (WGCNA)
\cite{langfelder2008b} is a versatile framework to extract networks
from high-dimensional data. It is able to identify biologically
functional subgroups, called modules, under many
differing settings \cite{eijk2012, horvath2012, langfelder2016}.  When relating
this structured information to the outcome of interest, 
additional challenges are faced. 
We are interested in the subgroups of features
which are most central to the function of the modules.
The method has to be even more selective in differentiating background noise from true connections to be used to explain the interplay of  differing  molecular levels, like gene expression and DNA methylation.

With Netboost, we propose a procedure to reduce dimensions
within high-dimensional datasets. We put a specific emphasis on large subgroups of
features that show a shared effect. For this we aggregate subgroup
information before applying the primary analysis strategy. In
public domain examples we show that we are able to extract patient
relevant information from multiple high-dimensional measurement types.

The paper is organized as follows. Section~\nameref{sec:methods} outlines the newly developed Netboost and describes the implementation. In Section~\nameref{sec:examples}
Netboost is applied in two public domain datasets and its performance is compared with two other approaches for each application. Section~\nameref{sec:discussion}
outlines the limitations and the potential future direction of the proposed method to
conclude the paper.

\section*{Methods}\label{sec:methods}
\begin{figure}[h!]
  \centerline{\includegraphics[width=0.49\textwidth]{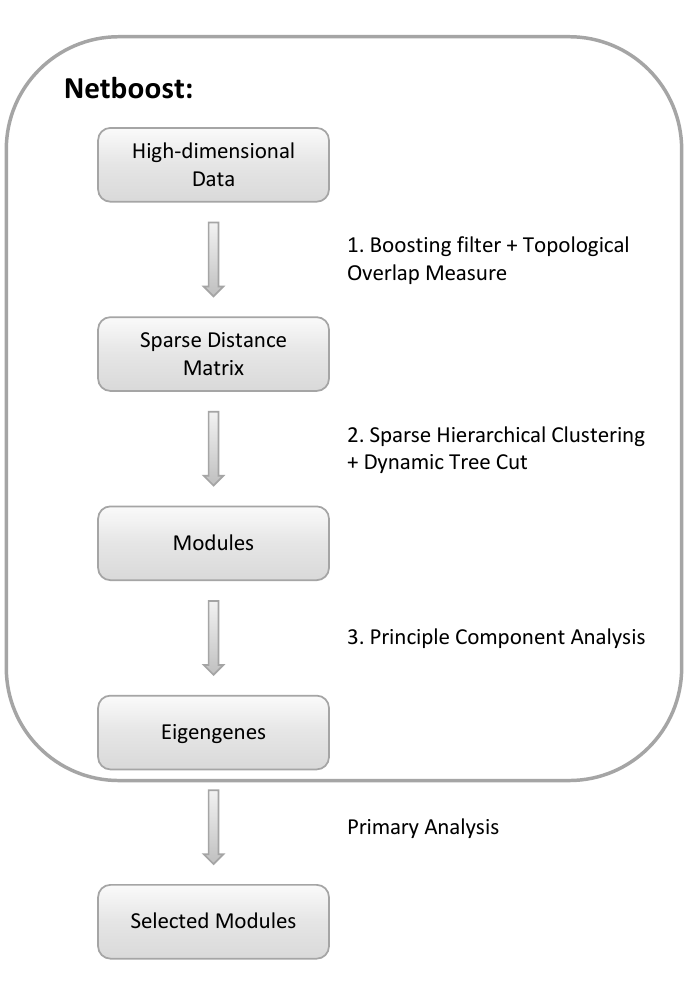}}
  \caption{\textbf{Netboost concept flow chart.}}\label{fig:concept}
\end{figure}

Netboost is a three-step procedure. As shown in
  Figure~\ref{fig:concept}, in the first step we calculate the boosting-based filter and a sparse distance matrix between features (Section~\nameref{subsec:boost} and \nameref{subsec:TOM}). From this combination we can reduce the network to its essential edges and remove spurious connections, originating from noise. We still retain the interconnectedness and stability of complex network structures including indirect connections that occur in many omics datasets reflecting biological pathway structures.
  
The second step consists of sparse hierarchical clustering and the dynamic tree cut procedure to determine modules from the dendrogram (Section~\nameref{subsec:upgma}) to transfer the network into a clustering.

Subsequently in step three, we aggregate the information in the
  modules by their first principal components (PCs) (Section~\nameref{sec:pc}) to achieve a low-dimensional representation of the original data.
 
 In this network-based
  dimension reduction method we modify the WGCNA by the addition of a
  multivariate filter and by the application of sparse hierarchical
  clustering.  
  Netboost is then followed by the primary analysis approach with the
  aggregated module information instead of with the original omics
  data. Here we present two applications, CoxBoost
  \cite{binder2008} to fit a Cox proportional hazards model
  integrated with a variable
  selection (Section~\nameref{sec:tcgaaml}) and a classification analysis by
  Random Forests \cite{breiman2001} (Section~\nameref{sec:hunt}).

\subsection*{Module detection}\label{sec:module}
Let $X$ be a $n\times p$-dimensional data-matrix, where $n\ll p$ with
$n$ being the number of samples and $p$ the number of features. We
assume $X$ to be continuous in each feature.

\subsubsection*{Boosting-based filter}\label{subsec:boost}
To first identify a general structure of our network we aggregate a
filter of important network edges by boosting.
Let $m \in \mathds{N}$ index samples, $i,j \in \mathds{N}$ index variables and 
\begin{align*}
  X_{i} := X_{m\leq n,i} && \text{and} && X_{-i} := X_{m\leq n,j \neq i}.
\end{align*}  
We fit
$$X_{i} = X_{-i}\beta_{j \leq p-1},$$
with $\beta \in \mathds{R}^{(p-1)}$.
Here, we perform component-wise likelihood-based boosting to fit a linear approximation of the outcome variable. In each iteration we fit the linear base learners using Fisher scoring with 
respect to the overall likelihood function one-by-one while keeping all other base learners fixed \cite{Tutz2006,Hastie1990}. Boosting is stopped after a given fixed number of steps. For all $\beta_j\neq 0$ we then add the tuple $(i,j)$ to the
filter. We estimate an undirected network, therefore the order of
$(i,j)$ is irrelevant. We define the filter by
$$\mathcal{F}:= \{(i,j),(j,i)|\exists i,j \in \mathds{N} \ \beta_i\neq 0 \text{ in }X_{j} = X_{-j}\beta \}.$$
By pruning the network to $\mathcal{F}$ we remove uninformative edges, reduce computational load and noise in subsequent steps.

\subsubsection*{Distance calculation}\label{subsec:TOM}
For tuples in $\mathcal{F}$ we define the adjacency of two features
analogous to \cite{langfelder2008b} by the power adjacency
function. For all other tuples the adjacency is set to 0. Hence, we have
$$
a_{ij} := \begin{cases} |\text{cor}(X_{i},X_{j})|^b & \text{if }
  (i,j)\in \mathcal{F}, \\
  0 &\text{else,}
\end{cases}
$$
where $b$ is chosen data-based by the scale free topology criterion
\cite{zhang2005} on a random subset of features. cor denotes the Pearson correlation.

We combine the topological overlap measure (TOM) \cite{ravasz2002} with $\mathcal{F}$ and
define

{\hspace*{-1.1cm}\vbox{\begin{align}\label{eq:TOM}
\text{TOM}_{ij}:=\begin{cases} 1 & \text{if } i = j, \\
 \frac{\sum_{u\neq i,j} a_{iu}a_{uj} + a_{ij}}{\min( \sum_{u\neq i} a_{iu}, \sum_{u\neq j} a_{uj})+1 - a_{ij}} & \text{if } (i,j)\in \mathcal{F}, \\
    0 &\text{else.}
  \end{cases}
\end{align}}}

As $a_{ij} \in [0,1]$ it follows that $\text{TOM}_{ij} \in
[0,1]$. These similarities are inverted to distances by
\begin{align}\label{eq:DistTOM}
  \text{DistTOM}_{ij}:=1-\text{TOM}_{ij}.
\end{align}

\subsubsection*{Hierarchical clustering and decomposition into modules}\label{subsec:upgma}
We apply the unweighted pair group method
with arithmetic mean (UPGMA) \cite{Sokal1958} to DistTOM. 
Parts of the network where no path exists in $\mathcal{F}$ are clustered independently.
A path between $X_{i}$ and $X_{j}$ exists exactly then when there is an $l \in \mathds{N}$ such that there are 
$t_{1\ldots l} \in \mathcal{F}$ with $i = t_{11}$,  $j = t_{l2}$ and
$\forall s: 1 \leq s \leq l-1 \ t_{s2}  = t_{(s+1)1}$. 
The dendrograms resulting from
these hierarchical clusterings are
separated into modules by the Dynamic Tree Cut procedure
\cite{langfelder2008}. Thus, features which are topologically
close on the filtered edges are grouped into modules. 

\subsection*{Aggregation of module information}\label{sec:pc}
By design the first PC explains the variation in one dimensional space as much as possible. Therefore, we aggregate the information in each module by its first PC, the so called eigengenes
\cite{langfelder2008b}. In a final step modules with highly correlated first principal components are merged to further reduce dimensionality. We define $E_q$ as the first principal component of
the $q$th module and
\begin{align*}
  X_\text{modules} := (E_1\T, \ldots, E_m\T),
\end{align*}  
where $m$ is the number of detected modules. $X_\text{modules}$ has the
dimension $n \times m$ where $m \ll p$. Due to its definition a
substantial part of
variation in $X$ is conserved in $X_\text{modules}$. At the same time the dimensionality is considerably reduced.

\subsection*{Module selection and evaluation}
Variable selection is performed analog to the primary analysis approach, but it is done on the set of eigengenes rather than on the set of features.

\subsubsection*{CoxBoost}\label{sec:coxboost}
We apply CoxBoost \cite{binder2008} to integrate the potentially still high-dimensional $X_\text{modules}$ with clinical covariates and
survival data as the primary outcome by likelihood-based
boosting. Analysis is implemented with the CoxBoost R package \cite{binder2013}. The
stopping criterion is chosen by cross-validation and a Cox
proportional hazards model is fitted.

\subsubsection*{Prediction errors}
To evaluate the performance of CoxBoost models we used the peperr R
package \cite{porzelius2009} which implements .632+ prediction errors
based on subsamples without replacement as recommended in
\cite{binder2008b}.
In high-dimensional data constellations bootstrap samples with replacement
often lead to overly complex models. Therefore, subsamples without
replacement of 63.2\% of the samples, which is equal to the expected number
of unique observations in one bootstrap sample drawn with replacement,
are implemented. Variability of prediction error curves is displayed by the distribution of integrated prediction error curves of the subsamples. 

\subsubsection*{Random forests}\label{sec:randomforests}
We apply random forests as described in \cite{breiman2001} to classify samples
based on $X_\text{modules}$ to their disease severity classes. 
To adequately explore the space of possible trees, also for the most high-dimensional of the analyses, we grow 10,000 trees in each analysis.

\subsection*{Implementation}
Netboost is built as an R package. It has been tested under
Linux and macOS. A Windows implementation is currently
not planned due to compiler dependencies.

As depicted in algorithm~\ref{alg:netboost} we first calculate
$\mathcal{F}$. Under the assumption of continuous $X_j$ and after
scaling and centering each we
efficiently implement the likelihood based boosting. The subsequent calculation of the
adjacencies and the
TOM are performed exclusively on network edges
in $\mathcal{F}$. Then the sparse distance matrix is exported to {\it
  Sparse} UPGMA by \cite{loewenstein2008}.
Here all empty edges where the nodes are connected indirectly are
assumed to have the maximal distance in the network and completely unconnected
nodes of the network are processed separately in independent clusterings.
This
agrees with the described method as all connected nodes not in
$\mathcal{F}$ have the distance of 1. By applying the filter we therefore reduce the
memory load and computational burden massively as the filter is
smaller than the whole network by orders of magnitudes as demonstrated in the examples in section~\nameref{sec:examples}.
\begin{algorithm}\label{alg:netboost}
 \KwIn{$X$, steps, minModuleSize, MEDissThres}
 \KwResult{$X_\text{modules}$}
 $\mathcal{F}=\emptyset$\;
 \For{$j\leftarrow 1$ \KwTo $p$}{
   fit $X_{j} = X_{-j}\beta$\;
   $\mathcal{F} = \mathcal{F} \cup  \{(i,j)|\exists i \in
\mathds{N} \ \beta_i\neq 0 \}$\;     
}
$\mathcal{F} = \{(i,j)|(i,j)\in \mathcal{F} \lor (j,i)\in \mathcal{F}\}$\;
randomFeatures = X[,sample(n= min(10,000,ncol(X))]\;
scaleFreeTopologyCriterium(randomFeatures) $\rightarrow$ $b$\;
\For{$(i,j)\in \mathcal{F}$}{
$a_{ij}=  |\text{cor}(X_{i},X_{j})|^b$\;
}
\For{$(i,j)\in \mathcal{F}$}{
compute DistTOM$_{ij}$ := equation \eqref{eq:DistTOM}\;
}
sparseUPGMA(DistTOM) $\rightarrow$ dendrogram\;
cutreeDynamic(dendrogram, minModuleSize) $\rightarrow$ modules\;
\For{$q \in$ modules}{
compute first principal component $E_q$\;
}
\While{$\exists q,q'$ with cor($E_q,E_{q'}$) $>(1-\text{MEDissThres})$}{
merge($q$, $q'$)\;
compute first principal component of merged module\;
}
\caption{Netboost}
\end{algorithm}

The algorithm is freely available as a Bioconductor R package at
http://bioconductor.org/packages/release/bioc/html/netboost.html. All functionality of Netboost is
available from within R whereas substantial parts of the
algorithm are implemented in C++. {\it Sparse} UPGMA is part of the
standalone MC-UPGMA software (for details see
\cite{loewenstein2008}). It is distributed with the Netboost R package. For extraction of modules we applied
the WGCNA \cite{langfelder2008b} and dynamicTreeCut
\cite{langfelder2008} R packages. As an example for the computational
demand Netboost was run on a dataset with 180 samples and 413,169
features (for details see section~\nameref{sec:tcgaaml}). Applying 
two Xeon E5 2690v3 at 2.6GHz (2x12cores)
and 40 GB of RAM it took Netboost 13.94 hours to compute.

\section*{Results}\label{sec:examples}
We apply Netboost to two datasets. In section~\nameref{sec:tcgaaml} it is applied to
DNA methylation and gene expression data from The Cancer Genome Atlas (TCGA) AML cohort to
predict survival (see~\nameref{sec:coxboost}). In section~\nameref{sec:hunt} it is applied to RNA sequencing data
to classify (see~\nameref{sec:randomforests}) Huntington's disease severity in mice.

\subsection*{TCGA AML: Methylation and gene expression predictive of overall survival}\label{sec:tcgaaml}
We selected the 180 AML patients in the public domain TCGA database
for which overall survival data, methylome and
gene expression measurements were available. TCGA data was
already preprocessed and normalized. Methylation was quantified with
Illumina Infinium HumanMethylation450 BeadChip arrays and gene
expression by Affymetrix HG U133 Plus 2.0 arrays. Thereby 396,065
methylation and 17,104 gene expression measurements were incorporated.
 We compared the following three schemes each with and
 without a clinical score. The clinical score is a dichotomized version of the linear predictor of a
 Cox proportional hazards regression model \cite{Therneau2000} of age at diagnosis and cytogenetic risk
 group, assessed as low, intermediate or high. Baseline hazards were estimated in separate strata according to sex. The score was solely used to evaluate Netboosts performance. 

In the models with the clinical score it was set as mandatory and thereby unpenalized in CoxBoost.
Thereby, DNA methylation and gene expression information was only added in these models if they could improve the prediction on top of the clinical score.
\begin{enumerate}
\item Direct application: Application of CoxBoost on the full dataset $X$.
\item Blockwise modules: The same approach as in Netboost but with the
  module detection done by blockwise WGCNA.
\item Netboost: Module PCs are calculated as described in
  section~\nameref{sec:methods}. CoxBoost is applied to these.
\end{enumerate}
The blockwise modules approach was the initial inspiration for the
Netboost method. They coincide with one another apart from Netboosts added
boosting-based filter and that $X$ has to be separated into feature
subsets for WGCNA so that the whole correlation matrix on the individual
subsets can be computed. This is accomplished with k-means
clustering and later aggregation via correlated eigengenes \cite{langfelder2008b}.

CoxBoost was implemented in all analyses in R with the peperr package
\cite{porzelius2009}. We used 10-fold cross validation to estimate
the optimal stopping criterion on the interval from 0 to 100. We applied 200 resampling steps to estimate the 632+ prediction errors.

\begin{figure*}[h!]
  \centerline{\includegraphics[width=\textwidth]{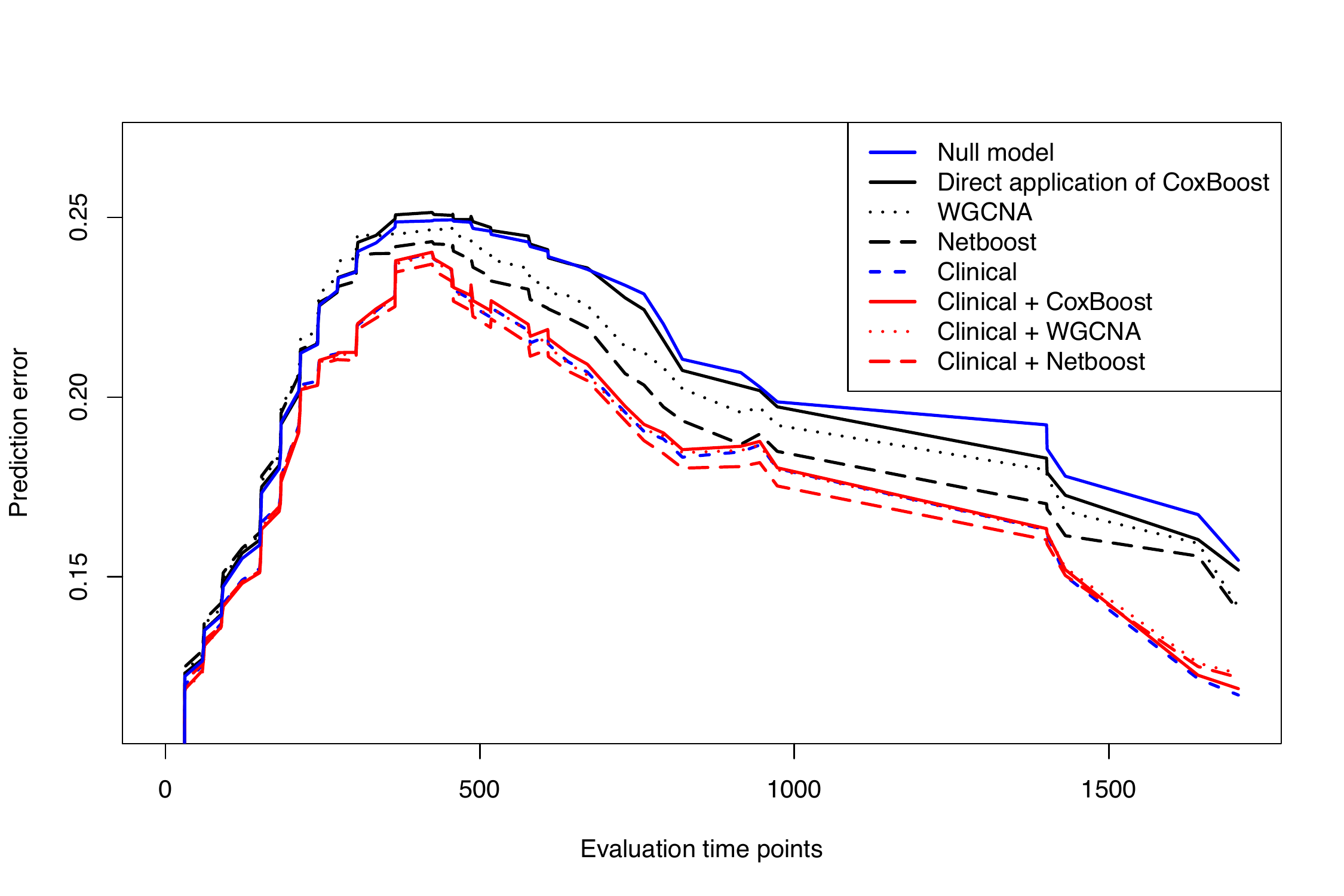}}
  \caption{\textbf{.632+ prediction error estimates for AML survival models.}}\label{fig:perr}
\end{figure*}

In models 1.-3. we integrated the analysis without the clinical score. The direct application on the full dataset, $X$, selected two
features and the 632+ prediction error curve, depicted in
Figure~\ref{fig:perr}, shows no improvement over the null model. The estimated .632+ prediction errors for days since diagnosis are
given in blue for the null model and dashed blue for the clinical
model. Prediction error curves based solely on DNA
methylation and gene expression are presented in black: The solid line for the
direct application of CoxBoost, the dotted line for the combination with
weighted gene co-expression network analysis (WGCNA) and the dashed line
for the combination with Netboost. The corresponding prediction
error curves additionally based on unpenalized clinical data are
presented in red.

Blockwise modules identified 568 modules with a mean module size of
671 in the range of 10 to 57,548. Ten was set as the minimum module
size. Henceforth, 92\% of the features were assigned to modules. The proportion of variance explained by eigengenes ranged from 23.9\% to 94.6\% (median = 50.5\%). In the WGCNA aggregated $X_\text{WGCNA modules}$ two modules were
selected by CoxBoost summarizing 26 features.

\begin{figure*}[h!]
  \centerline{\includegraphics[width=\textwidth]{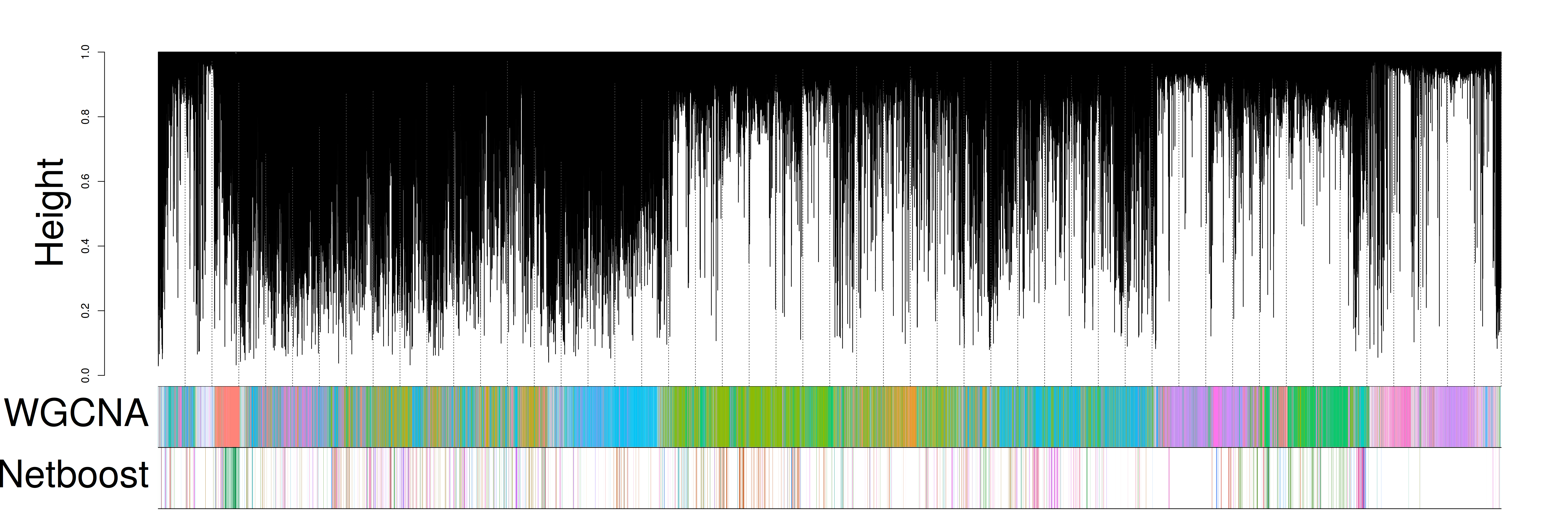}}
  \caption{\textbf{Dendrogram of the TCGA AML data.}
Dendrogram of the DNA methylation and gene expression features in the
TCGA AML data. The two rows below show the separation into modules by 
blockwise Weighted Gene Co-expression Network Analysis (WGCNA) and Netboost.}\label{fig:dendrogram}
\end{figure*}

For Netboost the multivariate filter was stopped after 20 steps and
resulted in a filter of 4,956,518 network edges. This
represents approximately $0.003\%$ of the edges. Based on
this Netboost identified 739 modules with an average module size of 52
in the range of 10 to 4,251. Accordingly 9\% of the features were
assigned to modules. The dendrogram based on the sparse network is depicted
in Figure~\ref{fig:dendrogram}. 
Netboost eigengenes generally explained a higher proportion of variance (median = 66.5\%, range = [45.7\%, 97.3\%]). 
CoxBoost selected six modules from the Netboost
aggregated $X_\text{Netboost  modules}$, summarizing 278
features. None of the features are shared by the selected Netboost modules and the selected
WGCNA modules.

\begin{figure}[h!]
  \centerline{\includegraphics[width=0.49\textwidth]{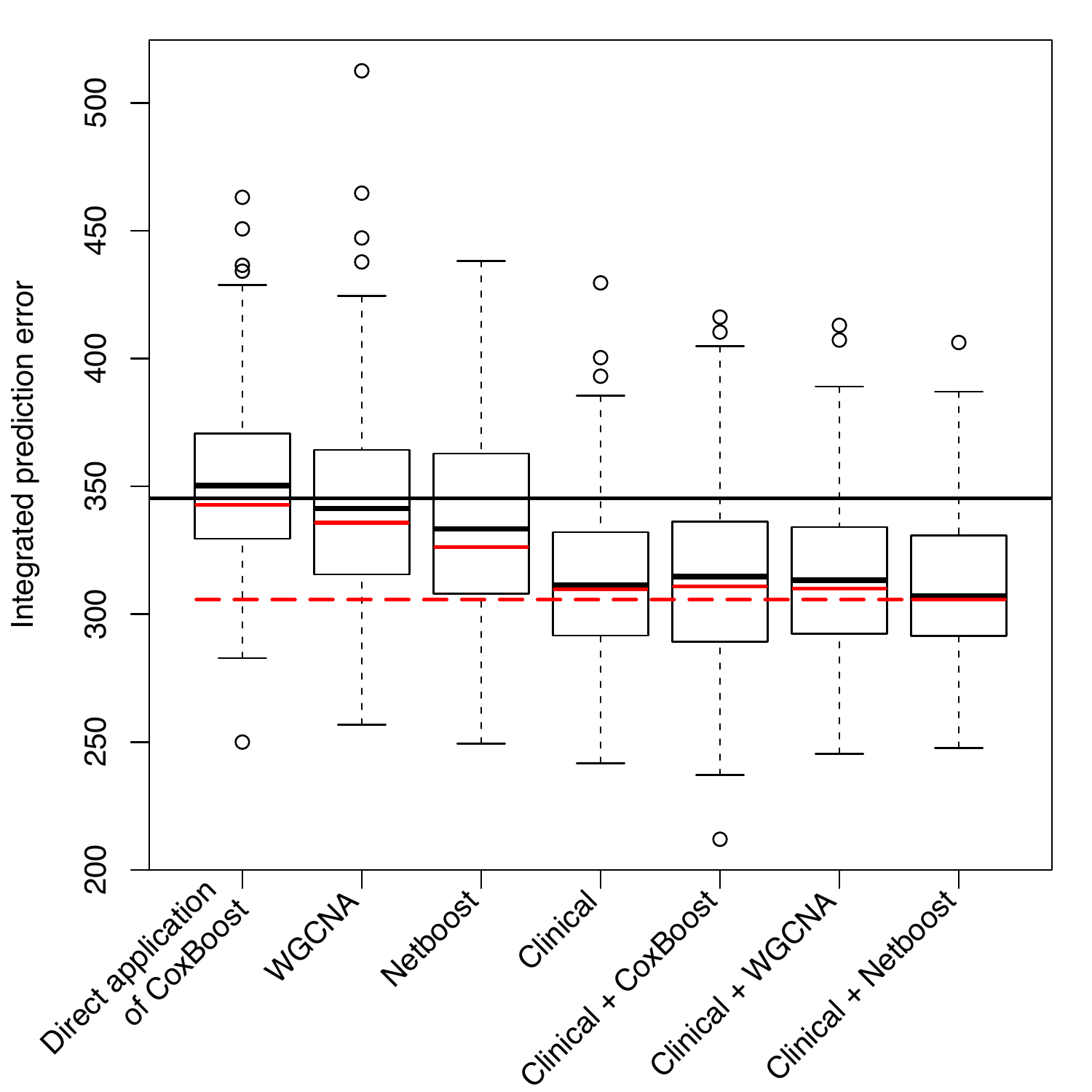}}
  \caption{\textbf{Variability of the .632+ prediction error
    estimates in AML survival models.}
  Integrated prediction error curve estimates from single subsamples for CoxBoost on the full dataset, CoxBoost on
  $X_\text{WGCNA}$ and CoxBoost on $X_\text{Netboost}$. The
  Kaplan-Meier benchmark value is indicated by a horizontal line. Red lines indicate the integrated .632+ prediction 
  error estimates with the line for the Clinical + Netboost model (lowest error) being extended by a dashed line.}\label{fig:iperr}
\end{figure}

As shown in Figure~\ref{fig:perr} the higher complexity indeed
corresponds to a better prediction performance in the 632+ prediction
errors. The blockwise modules approach was able to extract some
information but was outperformed by Netboost. This also
holds true when incorporating the variability of the individual 632+
resampling steps in Figure~\ref{fig:iperr}.

As depicted in Figure~\ref{fig:perr} and Figure~\ref{fig:iperr}, once we added
the clinical score as a mandatory covariate, none of the three
approaches was able to extract substantial additional information from the
molecular data. 
Overall, when comparing integrated prediction errors all analyses but the direct application of CoxBoost showed significant improvements over the 
null model (one-samples Student's t-test, p-value$<0.05$). Netboost including the clinical score had the lowest p-value (p-value = 1.3e-27). When comparing 
analyses with each other 
the integration with WGCNA and the Netboost significantly improved CoxBoost (p-value = 0.0437 and p-value = 0.0002, respectively) and Netboost improved the
accuracy of survival prediction on top of WGCNA (p-value = 0.0413). Furthermore, all analyses including the clinical score significantly improved prediction 
when compared with any analysis without the clinical score. In between analyses including the clinical score no significant differences were observed (two-samples Student's t-test, p-value$<0.05$).

\begin{figure}[h!]
  \centerline{\includegraphics[width=0.5\textwidth]{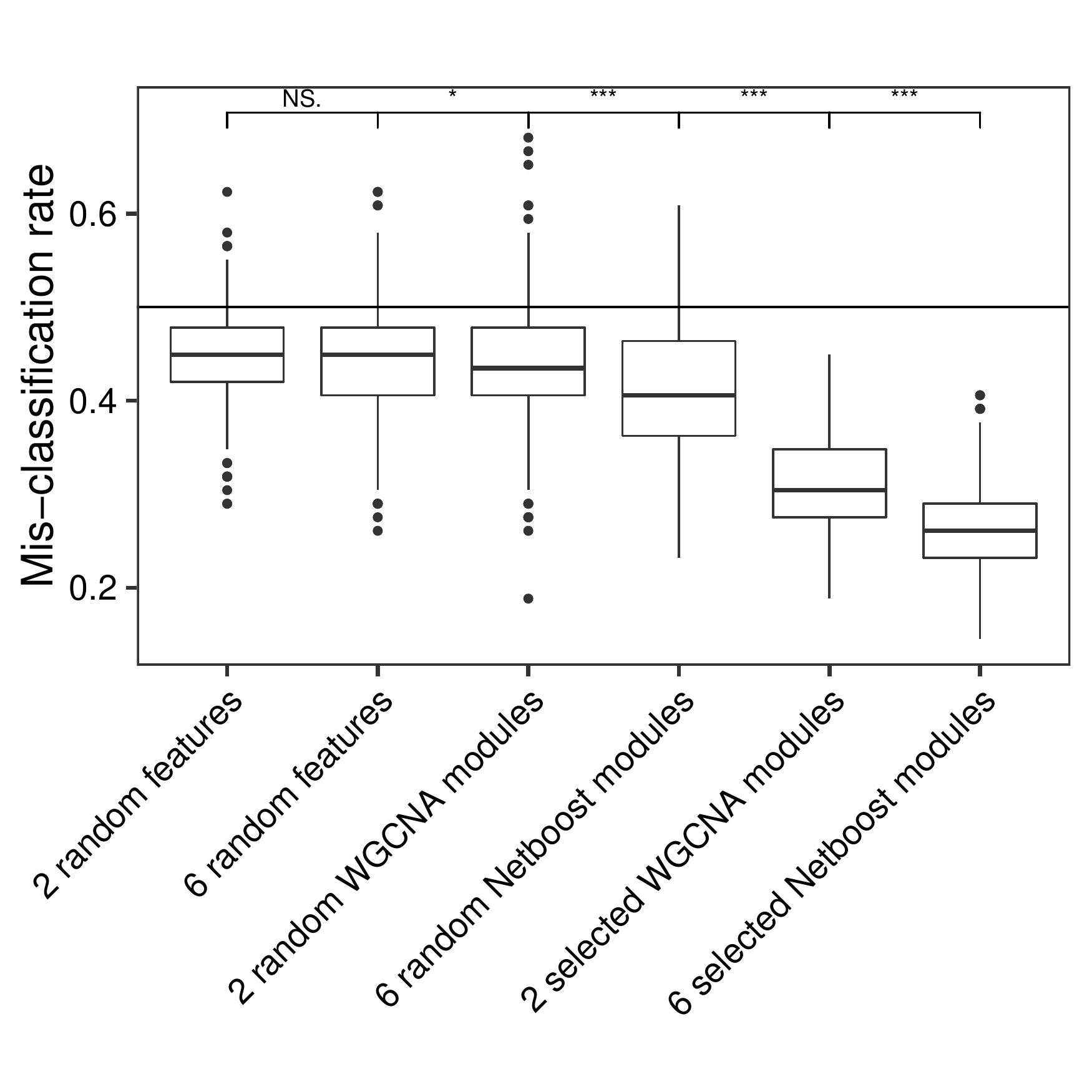}}
  \caption{\textbf{Mis-classification rate for logistic regression models of the clinical score in AML.}
    We compare randomly selected features of the raw data with randomly selected modules and the modules selected for survival prediction performance. The complexity of models is fixed to two and six to match the final survival models for Netboost and WGCNA respectively. The horizontal line indicates the expected mis-classification rate at random.        Asterisks indicate significance of unpaired two-samples Student's t-test (*** $p<$0.001, ** $p<$0.01, * $p<$0.05, NS. $p\geq$0.05). Only neighbouring columns were tested.
    }\label{fig:logreg}
\end{figure}

To investigate the possibility of the molecular
information extracted by Netboost being a surrogate for the clinical
score, we fitted logistic regression models for the module eigengenes
to the clinical score. We compared random selections of features out of all DNA methylation and gene expression features and modules, WGCNA and Netboost respectively, of similar size to the modules selected by WGCNA and Netboost and the modules selected for survival prediction. We fitted 500 models on subsamples of size 100 and evaluated the misclassification-rate on the remaining samples. For the random selections, features were reselected with each fit.
As shown in Figure~\ref{fig:logreg}, the
selected Netboost modules approximated the clinical score best.

\begin{figure*}[h!]
  \centerline{\includegraphics[width=\textwidth]{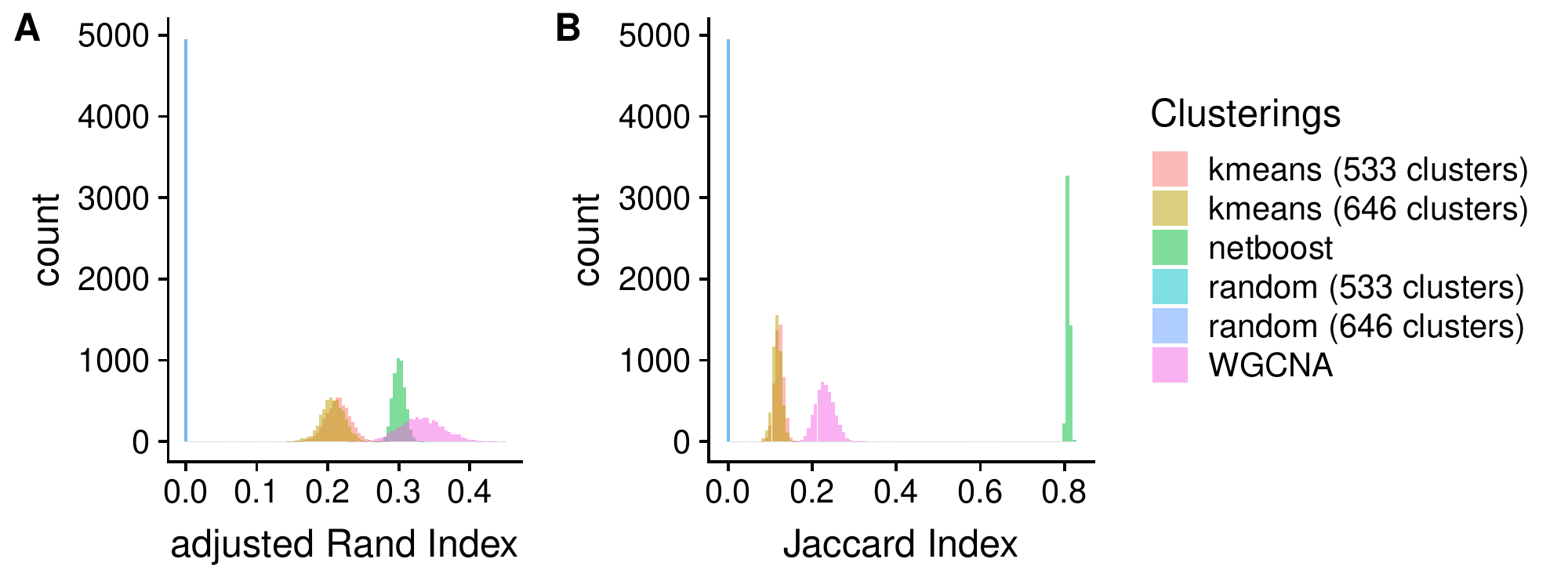}}
   \caption{\textbf{Clustering indices of the TCGA AML data.}
Histogram of cluster indices of 100 clusterings on random subsets of 100 samples applying Netboost, WGCNA, kmeans with $n$ classes and random selection of $n$ labels, where $n$ was set to the median number of modules in the Netboost runs (646) and WGCNA runs (533) . A) shows the adjusted Rand Index and B) the Jaccard Index.}\label{fig:cluster_indices}
\end{figure*}

To further comprehend the differences in the clustering we took random subsets of size 100 and compared the resulting Netboost and WGCNA clusterings using
 pairwise adjusted Rand Indices and Jaccard Indices \cite{wagner2007,Luxburg2010}. Additionally, we calculated kmeans clusterings with the number of clusters fixed to the median 
 number of clusters in Netboost clusterings (646) and WGCNA clusterings (533) and generated random clusterings with the respective number of clusters. 
The Rand Index is defined as the proportion of consistently clustered features between the two clusterings so tuples of features that are in both clusterings 
either in a common cluster or in both clusterings in differing clusters. The adjusted Rand Index corrects this for the expected number of consistent tuples 
given that the number of features and the number of clusters such that $E[$adjusted Rand Index$]=0$. The Jaccard Index is similar to the Rand Index, 
however it disregards tuples for features that are in different clusters for both clusterings. Both indices are less than or equal 1 and exactly 1 for identical clusterings.
As seen in Figure~\ref{fig:cluster_indices} both random clusterings had consistently pairwise indices of 0 and both kmeans clusterings were outperformed by 
WGCNA and Netboost with respect to both metrics. With respect to the adjusted Rand Index Netboosts median was below WGCNAs median while the order 
of minima was vice versa. When comparing the Jaccard Indices, Netboost outperforms WGCNA and shows a higher similarity for all pairwise comparisons with 
respect to this measure.

\begin{figure*}[h!]
  \centerline{\includegraphics[width=\textwidth]{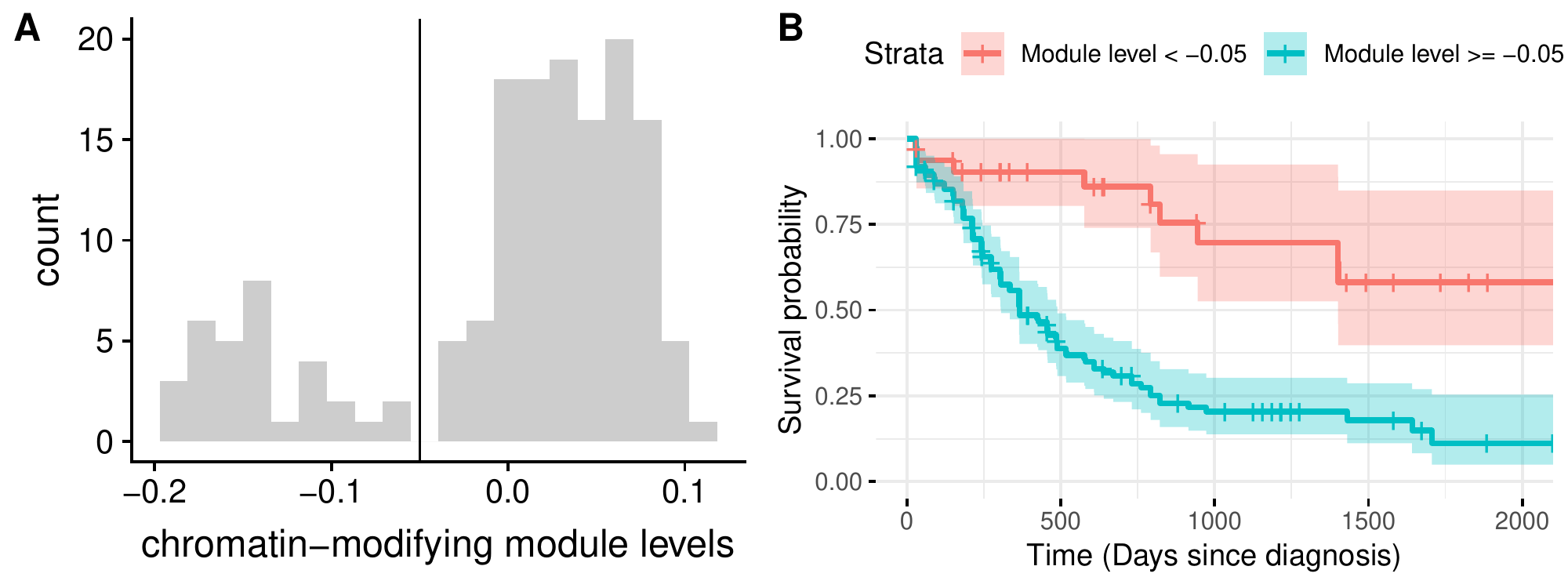}}
 \bigskip
  \centerline{\includegraphics[width=\textwidth]{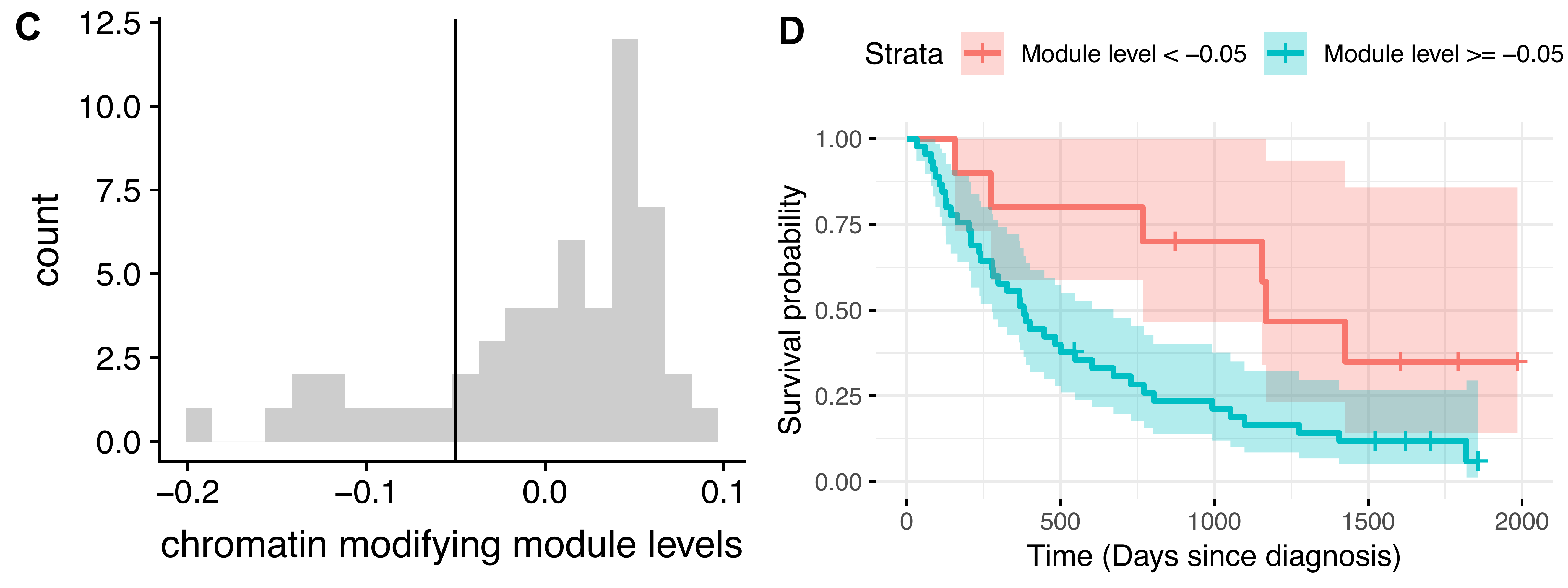}}
 \caption{\textbf{Chromatin modifying module eigengene distribution and Kaplan Meier in discovery and replication data.}
A) shows the bimodal eigengene distribution of the TCGA AML DNA methylation and gene expression module associated with chromatin modifying enzymes. The vertical line indicates at which point patients were stratified.
B) depicts the Kaplan Meier curves stratified by the modules eigengene for TCGA patients.
C) shows the bimodal eigengene distribution of the transferred module in AMLSG 12-09 DNA methylation data. The vertical line indicates at which point patients were stratified.
D) depicts the Kaplan Meier curves stratified by the modules eigengene for AMLSG 12-09 patients.
}\label{fig:ME71}
\end{figure*}

Netboost modules reflected known biology. 206 of the 739 Netboost modules consist of CpGs within 1,000 base pairs demonstrating the strength of local
 dependency in DNA methylation data. Netboost re-identified data-driven the association of CpGs in close proximity and \textit{cis} association of gene 
 methylation and expression.
In total, six different modules were selected that were variable in size and composition: 4 of the 6 modules consisted only of CpGs, one predominantly of CpGs
 and, in addition, 2 RNAs, one module only of 14 RNAs. The total number of CpGs varied from 10 to 88. The largest module (88 CpGs) contained numerous
  genes associated with hematopoiesis, such as {\it WT1} and {\it CXCL2}. The 2nd largest module (80 CpGs, 2 RNAs) represented 
  several genes encoding chromatin-modifying enzymes such as the H3K9 histone methyltransferase EHMT1 and the DNA demethylase TET3. To illustrate the strong association of this chromatin associated module alone we plotted stratified 
  Kaplan-Meier curves according to its bimodal distribution (Figure~\ref{fig:ME71}~A,B). The p-value of the likelihood ratio test of the dichotomised module levels (p-value = 7.0e-7) surpassed the one of the continues module levels (p-value = 4.0e-6); indicating that there might indeed be two states of these genes. Several 
  of these have already been implicated in AML pathogenesis and appear very promising for future predictive scores. Specifically, 4 CpGs mapped to the
   gene encoding EHMT1, also represented in the 4-gene methylation signature described by \cite{yamazaki2016}.

To validate the Netboost signature, we transferred it to DNA methylation data generated on
pre-treatment patient samples from the phase II AMLSG 12-09 study \cite{Schlenk2019}.
 In this study, DNA methylation based on the same Illumina Infinium 450k array and overall survival 
 was available for 55 AML patients. 
For processing and quality control of the raw methylation data, a customized version of the CPACOR
 pipeline \cite{Lehne2015} was used for data normalization and calculation of beta values. The
  complete preprocessing pipeline is available on Github (\url{https://github.com/genepi-freiburg/Infinium-preprocessing}). 
  As no data on gene expression was available one of the six modules could not be studied at all, while 2 were
   partially available (79 of 82 and 64 of 67 features) and 3 modules were available with all features. While the cox proportional 
    hazards model of these five modules was not significant in this smaller dataset (p-value = 0.4) the above mentioned chromatin 
    associated module alone did replicate (p-value = 0.04). Furthermore, this module exhibited a similar bimodal
     pattern as in TCGA and again, dichotomization led to a smaller p-value (p-value = 0.01, Figure~\ref{fig:ME71}~C,D).

\begin{figure*}[h!]
  \centerline{\includegraphics[width=\textwidth]{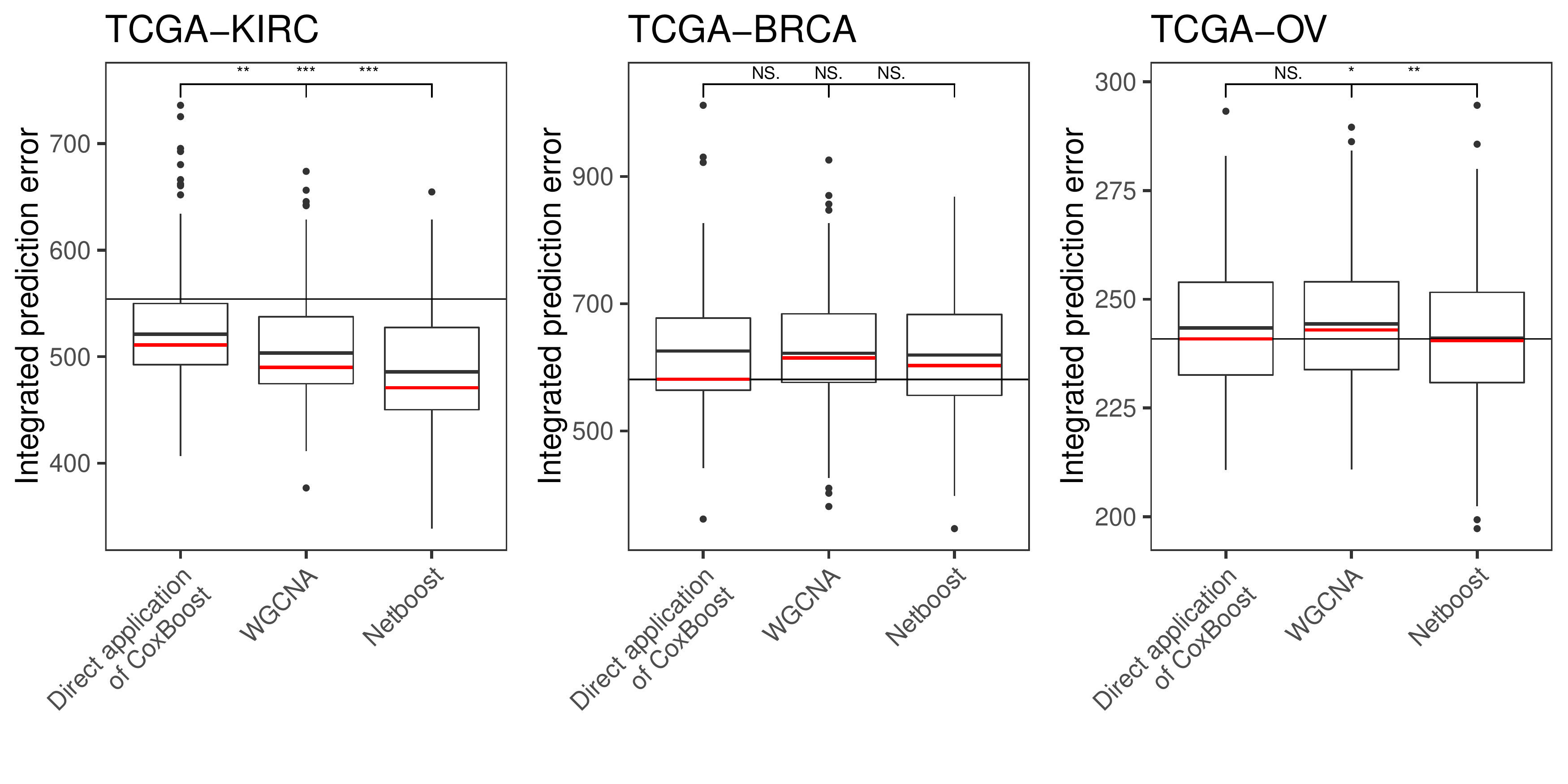}}
 \caption{\textbf{Variability of the .632+ prediction error
    estimates in TCGA Kidney Renal Clear Cell Carcinoma (KIRC), Breast Invasive Carcinoma (BRCA) and Ovarian Serous Cystadenocarcinoma (OV) survival models.}
  Integrated prediction error curve estimates from single subsamples for CoxBoost on the full dataset, CoxBoost on
  $X_\text{WGCNA}$ and CoxBoost on $X_\text{Netboost}$. The
  Kaplan-Meier benchmark value is indicated by a horizontal line.
  Red lines show the integrated .632+ prediction error estimates.
  Asterisks indicate significance of unpaired two-samples Student's t-test (*** $p<$0.001, ** $p<$0.01, * $p<$0.05, NS. $p\geq$0.05).
}\label{fig:intpred_other}
\end{figure*}

After the detailed analysis of the TCGA-AML DNA methylation and gene expression dataset we downloaded three more TCGA datasets; DNA methylation data of 774 breast invasive carcinoma (TCGA-BRCA) and 315 kidney renal clear cell carcinoma (TCGA-KIRC) patients and miRNA data of 464 ovarian serous cystadenocarcinoma (TCGA-OV) patients with available overall survival information. The 1,422 TCGA-OV miRNAs without missings and the 20,000 CpG sites with the largest variance for TCGA-BRCA and TCGA-KIRC respectively were selected for analysis. For each dataset we performed the same three analyses as for AML and calculated the 632+ prediction error estimates. Boxplots of the integrated prediction errors on the test set of the individual subsamplings are depicted in Figure~\ref{fig:intpred_other}. For clear cell carcinoma we observed similar performance as in AML. The integration with WGCNA significantly improved CoxBoost (p-value = 0.0013) and integration with Netboost improved the accuracy of survival prediction on top of WGCNA (p-value = 0.0006). For the other two datasets none of the three approaches was able to improve overall survival prediction.
 
\subsection*{Huntington's disease: Gene expression and CAG repeats}\label{sec:hunt}
Huntington's Disease (HD) is driven by the number of CAG repeats in the
huntingtin gene. In
\cite{langfelder2016} WGCNA revealed 13 striatal gene expression
modules that correlated with CAG length and age in a HD
knock-in mouse model. Further it was shown that several of
these effects translate to other HD models and patients. Recently, the analysis was extended to miRNA from the same mice in \cite{Langfelder2018}.

To evaluate the
performance of Netboost we used the mRNA dataset in an inverse
setup and determined the prediction errors in a classification
task. We selected the 48 mRNA-sequencing samples from mouse striatum
consisting of six
genetically engineered disease severities (20, 80, 92, 111, 140 and 175 CAG repeats) with four female and four
male mice all harvested at 6 months. We downloaded the preprocessed
mRNA-sequences from the
\href{https://www.ncbi.nlm.nih.gov/geo/query/acc.cgi?acc=GSE65774}{Gene
Expression Omnibus}. After removal of invariant transcripts, data
consisted of 28,010 transcripts.

We compared three setups:
\begin{enumerate}
\item Direct application: Random forest (RF) on the full dataset $X$.
\item Blockwise modules: Blockwise WGCNA + RF on module PCs
\item Netboost: RF on module PCs determined by Netboost
\end{enumerate}

RF was implemented in all analyses in R with the randomForest package
\cite{breiman2001}. We used 200 iterations of leave-one-out cross-validation.

The direct application on the full dataset, $X$, resulted in a mean
prediction error of $30.8\%$.

Blockwise modules identified 61 modules with a mean module size of 423
in the range of 11 to 6221. Ten was set as the
minimum module size. Henceforth, 92\% of the features were assigned to
modules. 
In the HD application the proportion of variance explained by eigengenes was lower than in the AML data 
(median = 42.1\%, range = [29.3\%, 63.4\%]).
On the WGCNA aggregated $X_\text{WGCNA modules}$ the mean
prediction error was $37.1\%$.

\begin{figure*}[h!]
  \centerline{\includegraphics[width=\textwidth]{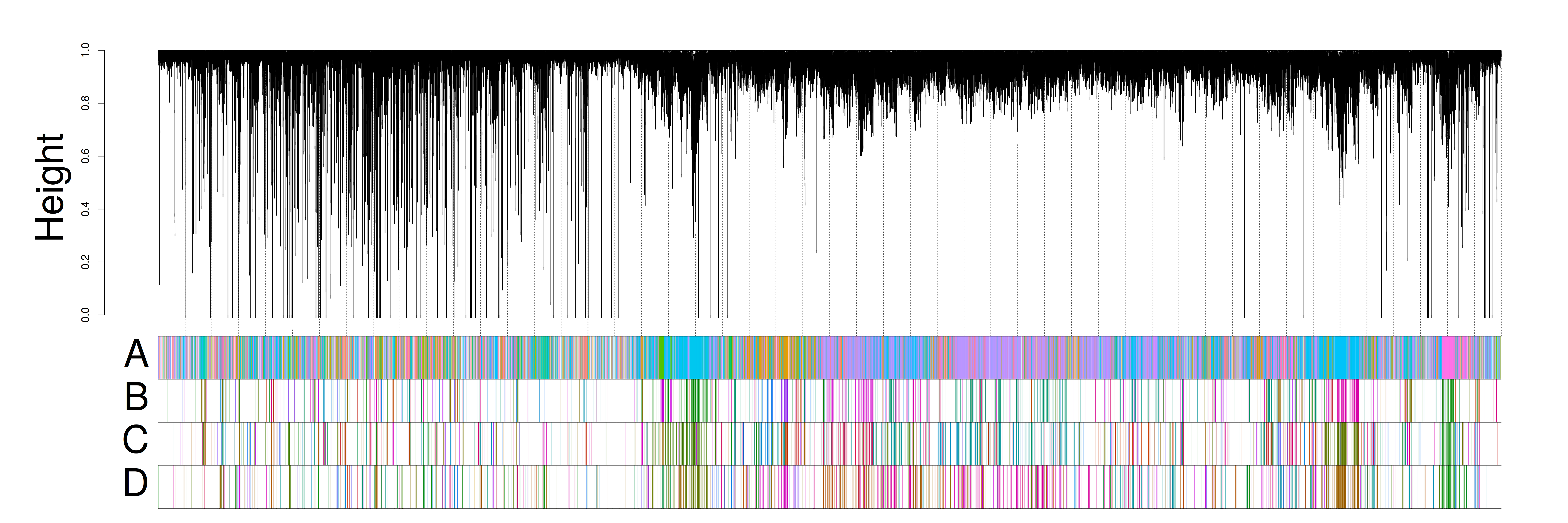}}
   \caption{\textbf{Dendrogram of Huntington's disease data.}
Dendrogram of the gene expression features in the
Huntington's disease data. A) shows the separation into modules by 
blockwise Weighted Gene Co-expression Network Analysis (WGCNA) and B), C) and D) show Netboost modules with 25, 20 and 15 boosting steps respectively.
}\label{fig:dendro_HD}
\end{figure*}

For Netboost the multivariate filter was stopped after 20 steps, resulting in a filter of 247,497 network edges. This represents
approximately $0.06\%$ of the edges. Based on this Netboost
identified 106 modules with an average module size of 46 in the range
of 10 to 561. Accordingly $17\%$ of the features were assigned to
modules.
Eigengenes of the Netboost modules explained a higher proportion of variance (median = 66.2\%, range = [52.3\%, 84.9\%]). 
On the Netboost aggregated $X_\text{Netboost modules}$ the mean
prediction error was $28.2\%$. The dendrogram based on the sparse network is depicted in Figure~\ref{fig:dendro_HD}. As shown for 25, 20 and 15 steps the clustering is stable under the choice of boosting steps.

Two-sample tests for equality of proportions with continuity correction showed significant differences in means of
prediction errors with Netboost errors being smaller than direct application
(p-value = 0.019) and WGCNA (p-value $<$ 2.2e-16) and direct application
errors being smaller than WGCNA (p-value $<$ 2.2e-16).

\section*{Discussion}\label{sec:discussion}
Netboost is designed in an unsupervised manner. While a supervised
approach to module detection would possibly further improve accuracy, these
approaches are complementary. 

In \cite{Reid2016} another approach for data with correlated
features is proposed. The first step is to cluster the features, and then choose
a cluster representative based on prediction performance. The second step is to apply either lasso or marginal significance testing on these
representatives. Apart from differences in the applied clustering
technique as with other supervised clustering techniques this might lead to
improved predictive performance but hinder the interpretation of the
selected clusters. 
Here, the primary aim is to maximize predictive power and thereby optimize the algorithm for biomarker detection.
This is done at the cost of potentially introducing some form of bias. 
Dependent on size and connectivity of the module the supervised selection might pick up peripheral features - voiding their function as a representative.

Thereby, we are hindered if we are interested in the biological interpretation of identified biomarkers in context of the network.
Keeping outcome and network
detection separate allows for unbiased interpretation of
any potential connections between subsequently in the primary analysis
selected modules and the outcome.

In \cite{Witten2009b} two extension to sparse canonical correlation analysis (CCA) \cite{Witten2009a} are introduced. First they propose a supervised form 
of sparse CCA and secondly
 they generalize the framework from two to multiple datasets. With this they offer a framework for identification of sparse linear combinations of the multiple sets of variables that are highly correlated with each other and associated with the outcome. While Netboost can also identify cross-omics correlations associated with the outcome, \cite{Witten2009b} omit within datatype connections and optimize their algorithm solely for cross-dataset combinations. 
 
 Starting with the WGCNA methodology our original design
 is the introduction of the filtering step before constructing the correlation-based network. Here, we chose a boosting based edge detection to allow for efficient 
 selection of essential edges. By introduction of this sparsity to the network we modified the TOM based distance and replaced UPGMA with the sparse UPGMA
  by Loewenstein et al \cite{loewenstein2008}. As Netboost is still based on the Pearson correlation coefficient and PCA based dimension reduction Netboost and 
  WGCNA share many of the same advantages and drawbacks due to their similarity in design.

Having a single representative for each
cluster might be an advantage for biomarker identification. Another
approach to consider for this purpose are hub genes replacing the eigengenes
we applied, as discussed in \cite{Yu2017} and
\cite{Das2017}. A hub gene is the most central node with the
highest connectivity of the module as
opposed to a summary measure and therefore allows cost efficient replication and
application as a biomarker \cite{Horvath2008}. Eigengenes might be superior in mechanistic studies, exploratory studies and the identification of previously unknown biological features. 

In general for dimension reduction of modules, eigengenes optimize explained variance with respect to the predefined dimensionality.
In our applications we fixed this to one to achieve comparability to WGCNA. However, our R package is more flexible than WGCNA and 
allows for the optional export of the first $i$ PCs with a fixed $i$ or for each module the first $j$ PCs which cumulatively explain at least $xy\%$ of variance.
Principal component based dimension reduction works particularly well if variables have linear relationships. If variables in modules would have non-linear relationships
other more flexible dimension reduction techniques such as autoencoders \cite{Bourlard1988,Kingma2013} might be more suited for calculation of aggregate measures. 
As the proposed module detection is ultimately based on the Pearson correlation coefficient which measures the strength of the linear relationship PCA based 
aggregation of modules is well defined as is observed by the high proportion of variance explained. If non-linear relationships between the features are of interest 
a complementary method would be required.

In Netboost feature-wise distances are defined based on Pearson
correlation coefficients, e.g. \cite{Yu2017} constructs networks
based on partial correlations. In the form of Gaussian graphical
models (GGM) partial correlations are frequently applied for
network construction
\cite{Krumsiek2011,Xie2016}. In \cite{Lee2017} GGMs are combined
with a filtering step to exclude insignificant edges from the network
much like Netboost. Partial correlations adjust for other
variables in the network and identify the independent connections
between features. In contrast, in Netboost we integrate indirect connections even further by the TOM. This is done to identify
interacting subgroups irrespective of whether this interaction is
direct or indirect. The focus lies on modules rather than on the individual edges, and the incorporation of indirect connections further stabilizes module detection.

As with GGMs a prime area of application beyond gene expression and DNA methylation is metabolome and proteome data. Due to their inherent co-regulation structures they offer themself to a network based analysis as was recently successfully demonstrated by combination of WGCNA  on proteome data and a subsequent genome wide association study in \cite{Emilsson2018}. 

Datatype specific features, like dependency of CpG sites in close proximity, are not incorporated as a-priori information in Netboost.
While this could inform the network, we prefer a universal design for omics data in general.
Therefore, a known biological nexus can be used for module evaluation as was done in Section~\nameref{sec:tcgaaml}.

In \cite{Zhang2013} the authors introduce Net-Cox which also
introduces network theory to improve survival prediction in a
high-dimensional context. In contrast to our combination of Netboost
and Coxboost, they introduce the estimated gene co-expression
structure directly to the penalty term of the Cox model. Net-Cox is
thereby inherently designed for survival analysis, whereas Netboost is
more flexible in its application.

In the section~\nameref{sec:hunt} WGCNA might be improved by the tuning of parameters as we only
applied the standard settings and the achieved clustering
superimposes the disease classifying signals. Netboost kept a more
compartmentalized and detailed network with the standard setting due
to the applied filtering step. 

In the shown applications we prefer specificity over sensitivity with respect to the clusterings. 
While it might be okay to miss an additional feature being part of a module we want to be sure about the selected features. 
Consistent with this, we deem the Jaccard Index as more important to our applications as most features are unrelated. 
As shown in Figure~\ref{fig:cluster_indices} Netboost is more robust than any of the competing clustering algorithms, when
 compared to WGCNA by the adjusted Rand Index and the Jaccard Index. To an extent it trades some sensitivity for specificity as expected 
 by the concept of integrating a filter on the network edges. With the adjusted Rand Index and Jaccard Index we chose basic measures of stability, which are especially reliable as they are used in a comparative fashion in identical resampling setting with different methodologies. When trying to assess stability of multi-level clustering structures \cite{Bertoni2008} or specific modules \cite{Langfelder2011} other measures might be more suitable. 

The applied clustering technique on the dendrogram in 
WGCNA and Netboost was introduced and compared to other methods in \cite{langfelder2008}. We kept the same cutting method to improve comparability.
Primary applications of WGCNA are related to identification of the network
structure itself. With respect to this we regard Netboost as a complementary 
approach designed to improve analysis including feature selection.

Netboost introduces the number of boosting steps as a parameter. This number can be chosen high as overfitting in filter estimation would only result in a less stringent filter rather than bias. Possible extensions include a probing based stopping criterion in the
boosting step. Boosting could be stopped by inserting uninformative features without the need to perform cross validation or a fixed
number of boosting steps during generation of the filter. This would automate the choice of boosting steps, while circumventing the often extensive additional computational load of cross-validation.
Another
direction to extend Netboost is inclusion of unclustered features which are currently ignored in primary analysis. This implies that isolated singular
features can not achieve a significant impact on the outcome. This is
not true for all settings. In the primary analysis $X_{\text{modules}}$
could be combined with a filtering method on the unclustered features.

\section*{Conclusion}
With Netboost we present an efficient dimension reduction technique
based on boosting and weighted gene co-expression networks distributed
as a Bioconductor R package. By introducing the boosting-based filter combined
with the TOM and sparse hierarchical
clustering combined with the dynamic tree cut procedure we were able
to extend efficiency and predictive performance simultaneously.

In the gene expression and DNA methylation examples this resulted in a
559- (TCGA AML) and 264-fold (HD) reduction of
features for the primary analyses. Choosing eigengenes as summary
measures we maximized the explained variance within each modules
without an assumptions-based extension to keep Netboost applicable to
a diverse set of biological experiments and primary analysis
strategies. Here, we displayed applications to {\it in vivo} DNA
methylation array, RNA array and RNA-seq measurements from patient and
mouse samples.
Paired with the clustering reflecting biological structures this leads
to improvements in highdimensional survival analysis as well as in
highdimensional classification.

In the Section~\nameref{sec:tcgaaml} the molecular prediction was improved by
identification of a surrogate for clinical information within the molecular
data and by the identification of hematopoietic genes and genes encoding chromatin-modifying enzymes.
In this application we were able to first abstract new features from the highdimensional data (modules), 
demonstrate a higher robustness than state-of-the-art alternative methods (cross-validation prediction errors) 
and validate the discovered correlates in an independent dataset (phase II AMLSG 12-09 study). 
Numerous of these genes have been suspected to play a role in AML pathogenesis before \cite{Rampal2014,yamazaki2016}.
Overall, the identified signature is promising for future research regarding AML pathogenesis and as a prognostic/predictive marker.
Furthermore, the association with chromatin-modifying enzymes could be replicated in an independent
  DNA methylation data set from the phase II AMLSG 12-09 clinical trial \cite{Schlenk2019} despite no 
  available gene expression measurements. 
  In the AMLSG 12-09 study, the effect of substituting cytarabine 
  by the DNA methyltransferase inhibitor 5-azacitidine in AML induction therapy was studied. This trial 
  tested the hypothesis that 5-azacytidine might reduce failure rates of intensive induction therapy 
  particularly in AML patients with unfavorable genetic features. It is of interest that validation of
   the chromatin associated module was successful in this independent AML patient DNA methylation data 
   set although the distribution of genetic aberrations in patients treated within the AMLSG 12-09 trial
    differed considerably from AML patients of the TCGA data set. Particularly, patients with core-binding factor 
    AML, AML with mutated NPM1, and AML with FLT3 internal tandem duplication were excluded in this trial.

In the section~\nameref{sec:hunt} Netboost outperformed the two other
approaches and achieved the lowest prediction error. In direct comparison to WGCNA, Netboost kept more compartmentalized networks with eigengenes better reflecting their respective module and these eigengenes exhibiting stronger associations with variables of interest.
Additionally, separation between, network detection and association with the trait of
interest, allows for unbiased analysis and interpretation of the
obtained structural information.
Due to this, our biological understanding of these complex diseases and experiments might benefit from the increase in prediction accuracy and added information via the
extracted network.

\section*{List of abbreviations}
\begin{description}
\item[AML:] Acute Myeloid Leukemia
\item[CAG repeat:] Repeat of the trinucleotide Cytosine, Adenine, Guanine
\item[CpG:] 5'---Cytosine---phosphate---Guanine---3'
\item[DNA:] Deoxyribonucleic Acid
\item[GGM:] Gaussian Graphical Models
\item[HD:] Huntington's Disease
\item[miRNA:] micro RNA
\item[mRNA:] messenger RNA
\item[PC:] Principal Component
\item[RF:] Random Forest
\item[RNA:] Ribonucleic Acid
\item[TCGA:] The Cancer Genome Atlas
\item[TOM:] Topological overlap measure
\item[UPGMA:] Unweighted Pair Group Method with Arithmetic Mean
\item[WGCNA:] Weighted Gene Co-expression Network Analysis
\end{description}

\section*{Declarations}
\subsection*{Acknowledgements}
The authors thank all members of the German-Austrian AML Study Group (AMLSG) for their participation in this study and for providing patient samples.

The results published here are in part based upon data generated by the TCGA Research Network: \url{https://www.cancer.gov/tcga}.
\subsection*{Competing interests}
The authors declare that they have no competing interests.
\subsection*{Funding}
This work has been supported by the Deutsche Forschungsgemeinschaft
 (German Research Foundation)
[SPP1463: LU 429/7-1, LU 429/8-1 to ML; FOR 2674: LU 429/16-1 (A05)]
 and by the German Consortium for Translational Cancer Research (DKTK) [L637 to ML]
  and by the Deutsche Krebshilfe [DKH110530 to LB and RC].
\subsection*{Authors' contributions}
PS, RC, ML wrote the manuscript,
PS, MaS, CP, RC, ML analyzed data,
PS, HB, MS developed the method,
PS, JK developed the R package,
PS, RC, ML, MS interpreted results,
KD, LB supervised the AMLSG 12-09 study,
MS supervised the research project, and 
all authors read and approved the final manuscript.

\section*{Availability of data and materials}
The datasets supporting the conclusions of this article are available in the GDC Data Portal repository, TCGA-LAML data [https://portal.gdc.cancer.gov/projects/TCGA-LAML] and the Gene Expression Omnibus [GSE65776, http://www.ncbi.nlm.nih.gov/geo/query/acc.cgi?acc=GSE65776].

\bibliographystyle{unsrt} 
\bibliography{Schlosser_et_al_2019} 

\begin{thebibliography}{10}

\bibitem{Saeys2007}
Yvan Saeys, Inaki Inza, and Pedro Larranaga.
\newblock A review of feature selection techniques in bioinformatics.
\newblock {\em Bioinformatics}, 23(19):2507--2517, 2007.

\bibitem{Sobieszkoda2017}
Dominika Sobieszkoda, Joanna Czech, Natalia Gablo, Marta Kopanska, Jacek
  Tabarkiewicz, Agnieszka Kolacinska, Tadeusz Robak, and Izabela Zawlik.
\newblock {MGMT} promoter methylation as a potential prognostic marker for
  acute leukemia.
\newblock {\em Archives of Medical Science : AMS}, 13(6):1433--1441, 10 2017.

\bibitem{Gardin2017}
Claude Gardin and Herv{\'e} Dombret.
\newblock {H}ypomethylating {A}gents as a {T}herapy for {AML}.
\newblock {\em Current Hematologic Malignancy Reports}, 12(1):1--10, Feb 2017.

\bibitem{Thomas2017}
Xavier Thomas and Caroline Le~Jeune.
\newblock {T}reatment of {E}lderly {P}atients {W}ith {A}cute {M}yeloid
  {L}eukemia.
\newblock {\em Current Treatment Options in Oncology}, 18(1):2, Jan 2017.

\bibitem{Papaemmanuil2016}
E.~Papaemmanuil, M.~Gerstung, L.~Bullinger, V.~I. Gaidzik, P.~Paschka, N.~D.
  Roberts, N.~E. Potter, M.~Heuser, F.~Thol, N.~Bolli, G.~Gundem, P.~Van~Loo,
  I.~Martincorena, P.~Ganly, L.~Mudie, S.~McLaren, S.~O'Meara, K.~Raine, D.~R.
  Jones, J.~W. Teague, A.~P. Butler, M.~F. Greaves, A.~Ganser, K.~Dohner, R.~F.
  Schlenk, H.~Dohner, and P.~J. Campbell.
\newblock {{G}enomic {C}lassification and {P}rognosis in {A}cute {M}yeloid
  {L}eukemia}.
\newblock {\em N. Engl. J. Med.}, 374(23):2209--2221, Jun 2016.

\bibitem{PradaArismendy2017}
Jeanette Prada-Arismendy, Johanna~C. Arroyave, and Sarah R{\"o}thlisberger.
\newblock {M}olecular biomarkers in acute myeloid leukemia.
\newblock {\em Blood Reviews}, 31(1):63 -- 76, 2017.

\bibitem{yamazaki2016}
J.~Yamazaki, R.~Taby, J.~Jelinek, N.~J. Raynal, M.~Cesaroni, S.~A. Pierce,
  S.~M. Kornblau, C.~E. Bueso-Ramos, F.~Ravandi, H.~M. Kantarjian, and J.~P.
  Issa.
\newblock {{H}ypomethylation of {T}{E}{T}2 {T}arget {G}enes {I}dentifies a
  {C}urable {S}ubset of {A}cute {M}yeloid {L}eukemia}.
\newblock {\em J. Natl. Cancer Inst.}, 108(2), Feb 2016.

\bibitem{meldi2015}
K.~Meldi, T.~Qin, F.~Buchi, N.~Droin, J.~Sotzen, J.~B. Micol,
  D.~Selimoglu-Buet, E.~Masala, B.~Allione, D.~Gioia, A.~Poloni, M.~Lunghi,
  E.~Solary, O.~Abdel-Wahab, V.~Santini, and M.~E. Figueroa.
\newblock {{S}pecific molecular signatures predict decitabine response in
  chronic myelomonocytic leukemia}.
\newblock {\em J. Clin. Invest.}, 125(5):1857--1872, May 2015.

\bibitem{langfelder2008b}
Peter Langfelder and Steve Horvath.
\newblock {WGCNA}: an {R} package for weighted correlation network analysis.
\newblock {\em BMC Bioinformatics}, page 559, 2008.

\bibitem{eijk2012}
K.~R. van Eijk, S.~de~Jong, M.~P. Boks, T.~Langeveld, F.~Colas, J.~H. Veldink,
  C.~G. de~Kovel, E.~Janson, E.~Strengman, P.~Langfelder, R.~S. Kahn, L.~H.
  van~den Berg, S.~Horvath, and R.~A. Ophoff.
\newblock {{G}enetic analysis of {D}{N}{A} methylation and gene expression
  levels in whole blood of healthy human subjects}.
\newblock {\em BMC Genomics}, 13:636, 2012.

\bibitem{horvath2012}
S.~Horvath, Y.~Zhang, P.~Langfelder, R.~S. Kahn, M.~P. Boks, K.~van Eijk, L.~H.
  van~den Berg, and R.~A. Ophoff.
\newblock {{A}ging effects on {D}{N}{A} methylation modules in human brain and
  blood tissue}.
\newblock {\em Genome Biol.}, 13(10):R97, 2012.

\bibitem{langfelder2016}
P.~Langfelder, J.~P. Cantle, D.~Chatzopoulou, N.~Wang, F.~Gao, I.~Al-Ramahi,
  X.~H. Lu, E.~M. Ramos, K.~El-Zein, Y.~Zhao, S.~Deverasetty, A.~Tebbe,
  C.~Schaab, D.~J. Lavery, D.~Howland, S.~Kwak, J.~Botas, J.~S. Aaronson,
  J.~Rosinski, G.~Coppola, S.~Horvath, and X.~W. Yang.
\newblock {{I}ntegrated genomics and proteomics define huntingtin {C}{A}{G}
  length-dependent networks in mice}.
\newblock {\em Nat. Neurosci.}, 19(4):623--633, Apr 2016.

\bibitem{binder2008}
H.~Binder and M.~Schumacher.
\newblock {{A}llowing for mandatory covariates in boosting estimation of sparse
  high-dimensional survival models}.
\newblock {\em BMC Bioinformatics}, 9:14, 2008.

\bibitem{breiman2001}
Leo Breiman.
\newblock {R}andom {F}orests.
\newblock {\em Machine Learning}, 45(1):5--32, Oct 2001.

\bibitem{Tutz2006}
G.~Tutz and H.~Binder.
\newblock {{G}eneralized additive modeling with implicit variable selection by
  likelihood-based boosting}.
\newblock {\em Biometrics}, 62(4):961--971, Dec 2006.

\bibitem{Hastie1990}
T.J. Hastie and R.J. Tibshirani.
\newblock {\em {G}eneralized {A}dditive {M}odels}.
\newblock Chapman \& Hall/CRC Monographs on Statistics \& Applied Probability.
  Taylor \& Francis, ~, 1990.

\bibitem{zhang2005}
Bin Zhang and Steve Horvath.
\newblock A general framework for weighted gene co-expression network analysis.
\newblock {\em Statistical Applications in Genetics and Molecular Biology}, 4,
  2005.

\bibitem{ravasz2002}
E.~Ravasz, A.~L. Somera, D.~A. Mongru, Z.~N. Oltvai, and A.~L. Barabasi.
\newblock {{H}ierarchical organization of modularity in metabolic networks}.
\newblock {\em Science}, 297(5586):1551--1555, Aug 2002.

\bibitem{Sokal1958}
R.R. Sokal, C.D. Michener, and University of~Kansas.
\newblock {\em A Statistical Method for Evaluating Systematic Relationships}.
\newblock University of Kansas science bulletin. University of Kansas, ~, 1958.

\bibitem{langfelder2008}
Peter Langfelder, Bin Zhang, and Steve Horvath.
\newblock Defining clusters from a hierarchical cluster tree: the {Dynamic}
  {Tree} {Cut} package for {R}.
\newblock {\em Bioinformatics}, 24:719--720, 2008.

\bibitem{binder2013}
Harald Binder.
\newblock {\em CoxBoost: Cox models by likelihood based boosting for a single
  survival endpoint or competing risks}, 2013.
\newblock R package version 1.4.

\bibitem{porzelius2009}
C.~Porzelius, H.~Binder, and M.~Schumacher.
\newblock {{P}arallelized prediction error estimation for evaluation of
  high-dimensional models}.
\newblock {\em Bioinformatics}, 25(6):827--829, Mar 2009.

\bibitem{binder2008b}
Harald Binder and Martin Schumacher.
\newblock {A}dapting {P}rediction {E}rror {E}stimates for {B}iased {C}omplexity
  {S}election in {H}igh-{D}imensional {B}ootstrap {S}amples.
\newblock {\em Statistical Applications in Genetics and Molecular Biology},
  7(1), 01 2008.

\bibitem{loewenstein2008}
Yaniv Loewenstein, Elon Portugaly, Menachem Fromer, and Michal Linial.
\newblock Efficient algorithms for accurate hierarchical clustering of huge
  datasets: tackling the entire protein space.
\newblock {\em Bioinformatics}, 24(13):41--49, 2008.

\bibitem{Therneau2000}
T.M. Therneau and P.M. Grambsch.
\newblock {\em Modeling Survival Data: Extending the Cox Model}.
\newblock Statistics for Biology and Health. Springer, ~, 2000.

\bibitem{wagner2007}
Silke Wagner and Dorothea Wagner.
\newblock Comparing clusterings - an overview.
\newblock Technical Report~4, {Karlsruhe}, 2007.

\bibitem{Luxburg2010}
Ulrike von Luxburg.
\newblock Clustering stability: An overview.
\newblock {\em Foundations and Trends in Machine Learning}, 2(3):235--274,
  2010.

\bibitem{Schlenk2019}
R.~F. Schlenk, D.~Weber, W.~Herr, G.~Wulf, H.~R. Salih, H.~G. Derigs,
  A.~Kuendgen, M.~Ringhoffer, B.~Hertenstein, U.~M. Martens, M.~Griesshammer,
  H.~Bernhard, J.~Krauter, M.~Girschikofsky, D.~Wolf, E.~Lange, J.~Westermann,
  E.~Koller, S.~Kremers, M.~Wattad, M.~Heuser, F.~Thol, G.~Gohring, D.~Haase,
  V.~Teleanu, V.~Gaidzik, A.~Benner, K.~D{\"o}hner, A.~Ganser, P.~Paschka, and
  H.~D{\"o}hner.
\newblock {{R}andomized phase-{I}{I} trial evaluating induction therapy with
  idarubicin and etoposide plus sequential or concurrent azacitidine and
  maintenance therapy with azacitidine}.
\newblock {\em Leukemia}, Feb 2019.

\bibitem{Lehne2015}
B.~Lehne, A.~W. Drong, M.~Loh, W.~Zhang, W.~R. Scott, S.~T. Tan, U.~Afzal,
  J.~Scott, M.~R. Jarvelin, P.~Elliott, M.~I. McCarthy, J.~S. Kooner, and J.~C.
  Chambers.
\newblock {{A} coherent approach for analysis of the {I}llumina
  {H}uman{M}ethylation450 {B}ead{C}hip improves data quality and performance in
  epigenome-wide association studies}.
\newblock {\em Genome Biol.}, 16:37, Feb 2015.

\bibitem{Langfelder2018}
Peter Langfelder, Fuying Gao, Nan Wang, David Howland, Seung Kwak, Thomas~F.
  Vogt, Jeffrey~S. Aaronson, Jim Rosinski, Giovanni Coppola, Steve Horvath, and
  X.~William Yang.
\newblock {M}icrorna signatures of endogenous huntingtin cag repeat expansion
  in mice.
\newblock {\em PLOS ONE}, 13(1):1--20, 01 2018.

\bibitem{Reid2016}
Stephen Reid and Robert Tibshirani.
\newblock {S}parse regression and marginal testing using cluster prototypes.
\newblock {\em Biostatistics}, 17(2):364--376, 2016.

\bibitem{Witten2009b}
D.~M. Witten and R.~J. Tibshirani.
\newblock {{E}xtensions of sparse canonical correlation analysis with
  applications to genomic data}.
\newblock {\em Stat Appl Genet Mol Biol}, 8:Article28, 2009.

\bibitem{Witten2009a}
D.~M. Witten, R.~Tibshirani, and T.~Hastie.
\newblock {{A} penalized matrix decomposition, with applications to sparse
  principal components and canonical correlation analysis}.
\newblock {\em Biostatistics}, 10(3):515--534, Jul 2009.

\bibitem{Yu2017}
Donghyeon Yu, Johan Lim, Xinlei Wang, Faming Liang, and Guanghua Xiao.
\newblock {E}nhanced construction of gene regulatory networks using hub gene
  information.
\newblock {\em BMC Bioinformatics}, 18(1):186, Mar 2017.

\bibitem{Das2017}
Samarendra Das, Prabina~Kumar Meher, Anil Rai, Lal~Mohan Bhar, and Baidya~Nath
  Mandal.
\newblock {S}tatistical {A}pproaches for {G}ene {S}election, {H}ub {G}ene
  {I}dentification and {M}odule {I}nteraction in {G}ene {C}o-{E}xpression
  {N}etwork {A}nalysis: {A}n {A}pplication to {A}luminum {S}tress in {S}oybean
  ({G}lycine max {L}.).
\newblock {\em PLOS ONE}, 12(1):1--24, 01 2017.

\bibitem{Horvath2008}
Steve Horvath and Jun Dong.
\newblock {G}eometric {I}nterpretation of {G}ene {C}oexpression {N}etwork
  {A}nalysis.
\newblock {\em PLOS Computational Biology}, 4(8):1--27, 08 2008.

\bibitem{Bourlard1988}
H.~Bourlard and Y.~Kamp.
\newblock {{A}uto-association by multilayer perceptrons and singular value
  decomposition}.
\newblock {\em Biol Cybern}, 59(4-5):291--294, 1988.

\bibitem{Kingma2013}
Diederik~P Kingma and Max Welling.
\newblock Auto-encoding variational bayes.
\newblock {\em arXiv preprint arXiv:1312.6114}, 2013.

\bibitem{Krumsiek2011}
Jan Krumsiek, Karsten Suhre, Thomas Illig, Jerzy Adamski, and Fabian~J. Theis.
\newblock {G}aussian graphical modeling reconstructs pathway reactions from
  high-throughput metabolomics data.
\newblock {\em BMC Systems Biology}, 5(1):21, Jan 2011.

\bibitem{Xie2016}
Yuying Xie, Yufeng Liu, and William Valdar.
\newblock {J}oint estimation of multiple dependent {G}aussian graphical models
  with applications to mouse genomics.
\newblock {\em Biometrika}, 103(3):493--511, 2016.

\bibitem{Lee2017}
Sangin Lee, Faming Liang, Ling Cai, and Guanghua Xiao.
\newblock Integrative analysis of gene networks and their application to lung
  adenocarcinoma studies.
\newblock {\em Cancer Informatics}, 16:1176935117690778, 2017.

\bibitem{Emilsson2018}
V.~Emilsson, M.~Ilkov, J.~R. Lamb, N.~Finkel, E.~F. Gudmundsson, R.~Pitts,
  H.~Hoover, V.~Gudmundsdottir, S.~R. Horman, T.~Aspelund, L.~Shu, V.~Trifonov,
  S.~Sigurdsson, A.~Manolescu, J.~Zhu, O.~Olafsson, J.~Jakobsdottir, S.~A.
  Lesley, J.~To, J.~Zhang, T.~B. Harris, L.~J. Launer, B.~Zhang,
  G.~Eiriksdottir, X.~Yang, A.~P. Orth, L.~L. Jennings, and V.~Gudnason.
\newblock {{C}o-regulatory networks of human serum proteins link genetics to
  disease}.
\newblock {\em Science}, 361(6404):769--773, 08 2018.

\bibitem{Zhang2013}
Wei Zhang, Takayo Ota, Viji Shridhar, Jeremy Chien, Baolin Wu, and Rui Kuang.
\newblock {N}etwork-based {S}urvival {A}nalysis {R}eveals {S}ubnetwork
  {S}ignatures for {P}redicting {O}utcomes of {O}varian {C}ancer {T}reatment.
\newblock {\em PLOS Computational Biology}, 9(3):1--16, 03 2013.

\bibitem{Bertoni2008}
A.~Bertoni and G.~Valentini.
\newblock {{D}iscovering multi-level structures in bio-molecular data through
  the {B}ernstein inequality}.
\newblock {\em BMC Bioinformatics}, 9 Suppl 2:S4, Mar 2008.

\bibitem{Langfelder2011}
P.~Langfelder, R.~Luo, M.~C. Oldham, and S.~Horvath.
\newblock {{I}s my network module preserved and reproducible?}
\newblock {\em PLoS Comput. Biol.}, 7(1):e1001057, Jan 2011.

\bibitem{Rampal2014}
R.~Rampal, A.~Alkalin, J.~Madzo, A.~Vasanthakumar, E.~Pronier, J.~Patel, Y.~Li,
  J.~Ahn, O.~Abdel-Wahab, A.~Shih, C.~Lu, P.~S. Ward, J.~J. Tsai, T.~Hricik,
  V.~Tosello, J.~E. Tallman, X.~Zhao, D.~Daniels, Q.~Dai, L.~Ciminio,
  I.~Aifantis, C.~He, F.~Fuks, M.~S. Tallman, A.~Ferrando, S.~Nimer,
  E.~Paietta, C.~B. Thompson, J.~D. Licht, C.~E. Mason, L.~A. Godley,
  A.~Melnick, M.~E. Figueroa, and R.~L. Levine.
\newblock {{D}{N}{A} hydroxymethylation profiling reveals that {W}{T}1
  mutations result in loss of {T}{E}{T}2 function in acute myeloid leukemia}.
\newblock {\em Cell Rep}, 9(5):1841--1855, Dec 2014.

\end{thebibliography}
\newpage

\section*{Figures}

\end{document}